\documentclass[conference]{IEEEtran}
\usepackage{amsmath, amssymb, graphicx, xcolor, multirow, multicol, comment, subcaption, algorithm2e, url, float, etoolbox, adjustbox, pgf, soul, geometry, colortbl, booktabs}

\apptocmd{\thebibliography}{\small}{}{}
\patchcmd{\thebibliography}{\leftmargin\labelwidth}{\leftmargin\labelwidth\itemsep=0pt\parsep=0pt\topsep=0pt}{}{}

\title{Representation Loss Minimization with Randomized Selection Strategy for Efficient Environmental Fake Audio Detection}

\author{
    \IEEEauthorblockN{Orchid Chetia Phukan\textsuperscript{*1}, Girish\textsuperscript{*1,2}, Mohd Mujtaba Akhtar\textsuperscript{*1}, Swarup Ranjan Behera\textsuperscript{*3}, \\ Nitin Choudhury\textsuperscript{1},
    Arun Balaji Buduru\textsuperscript{1}, Rajesh Sharma\textsuperscript{1,4}, S.R Mahadeva Prasanna\textsuperscript{5,6}}
    \IEEEauthorblockA{
        $^1$\textit{IIIT-Delhi, India},
        $^2$\textit{UPES, India},
        $^3$\textit{Reliance Jio AICoE, India},
        $^4$\textit{University of Tartu, Estonia} \\
        $^5$\textit{IIT-Dharwad, India},
        $^6$\textit{IIIT-Dharwad, India}\\
        *Equal contribution\\
        orchidp@iiitd.ac.in
    }
}

\begin{document}

\maketitle

\begin{abstract}
The adaptation of foundation models has significantly advanced environmental audio deepfake detection (EADD), a rapidly growing area of research. These models are typically fine-tuned or utilized in their frozen states for downstream tasks. However, the dimensionality of their representations can substantially lead to a high parameter count of downstream models, leading to higher computational demands. So, a general way is to compress these representations by leveraging state-of-the-art (SOTA) unsupervised dimensionality reduction techniques (PCA, SVD, KPCA, GRP) for efficient EADD. However, with the application of such techniques, we observe a drop in performance. So in this paper, we show that representation vectors contain redundant information, and randomly selecting 40-50\% of representation values and building downstream models on it preserves or sometimes even improves performance. We show that such random selection preserves more performance than the SOTA dimensionality reduction techniques while reducing model parameters and inference time by almost over half.

\end{abstract} 


\section{Introduction}
Imagine a scenario where a gunshot sound, convincingly fabricated by an environmental audio deepfake, 
is broadcast over a school’s public address system. This meticulously crafted audio, indistinguishable from a real gunshot, could incite mass panic, activate emergency protocols, and divert critical resources from genuine emergencies. Such situations illustrate the profound risks posed by synthetic environmental sounds to audio authentication and security. Unlike traditional deepfakes~\cite{DF1,DF2,DF3,DF4}, which primarily target human speech, these advanced models~\cite{ouajdi2024detection,choi2023foley} can generate complex soundscapes, like realistic gunshots, using sophisticated deep learning techniques. While synthetic environmental sounds have promising applications in virtual reality and immersive technologies, their potential for misuse raises serious ethical and security concerns, including misinformation, fraudulent audio evidence, and privacy breaches.

Recent advancements in deep learning have transformed synthetic audio detection, moving beyond classical ML approaches~\cite{OLD1,OLD2} to neural network-based models~\cite{DEEP1,DEEP2,DEEP3}. 
More recently, transformer-based foundation models (FMs) like Wav2Vec2 \cite{wang2021investigating}, MMS, Whisper \cite{chetia-phukan-etal-2024-heterogeneity} have shown SOTA performance in synthetic audio detection. These FMs provide performance benefits as well as training efficiency, preventing training models from scratch. These developments underscore the importance of FMs - very large neural networks pre-trained on vast datasets - which are fine-tuned or applied as feature extractors to downstream tasks. These FMs have also provided a substantial boost towards better environmental audio deepfake detection (EADD) \cite{ouajdi2024detection}, our main focus in this study. 

However, the high dimensionality of representation vectors from these FMs presents a significant challenge. While higher-dimensional representations often enhance accuracy, they also increase the parameter count in downstream models, leading to higher computational demands and more resource-intensive deployments. In general, to cut down such high computational demands, the best and easy-to-go approach would be to compress these high-dimension representation vectors with state-of-the-art (SOTA) dimensionality reduction techniques. However, the usage of such techniques often leads to a decrease in performance. To solve this, our research provides an innovative approach that starts with asking
two critical questions: (1) \textit{Do representation vectors contain redundant information?} and (2) \textit{Can we reduce the number of representation values without compromising performance?} 

\noindent The main contributions are as follows:

\begin{itemize} 
\item \textit{Dimensionality Reduction Optimization:} We show that randomly selecting 40-50\% of representation values of FMs maintains or even improves performance while cutting parameter counts and inference time by almost half, outperforming several SOTA dimensionality reduction techniques.

\item \textit{Broad Applicability:} To validate that random selection works and is transferable across different FMs, we perform a comprehensive empirical analysis across SOTA audio FMs and multimodal FMs. We show that such behavior exists across different FMs and demonstrating that full representation vectors are not always necessary, offering a practical and simple solution for efficient EADD. 
\end{itemize}

\noindent We will release the models and codes for the experiments carried out as part of the study after the review process. 

\section{Foundation Models}
In this section, we discuss the audio foundation models (AFMs) and multimodal foundation models (MFMs) used in our study. 

\subsection{Audio Foundation Models}

\noindent \textbf{Unispeech-SAT\footnote{\url{https://huggingface.co/microsoft/unispeech-sat-base}}  \cite{chen2022unispeech}:} It is trained through self-supervised learning by incorporating an utterance-wise contrastive loss and multi-task learning with a speaker-aware format. 

\noindent \textbf{WavLM\footnote{\url{https://huggingface.co/microsoft/wavlm-base}} \cite{chen2022wavlm}:} WavLM was proposed as a general-purposed FM for various speech processing tasks. It performs both masked speech prediction and speech denoising during pre-training. WavLM shows SOTA performance in SUPERB benchmark~\cite{Yang2021SUPERBSP}. 

\noindent \textbf{Wav2vec2\footnote{\url{https://huggingface.co/facebook/wav2vec2-base}} \cite{baevski2020wav2vec}:} It leverages self-supervised learning for pre-training and masks the input speech in latent space while solving a constrastive tasks. It outperforms previous models and achieves SOTA performance on Librispeech. 

\noindent \textbf{TRILLsson\footnote{\url{https://tfhub.dev/google/nonsemantic-speech-benchmark/trillsson4/1}} \cite{shor22_interspeech}:} It is distilled version of the universal paralinguistic conformer. TRILLsson also shows SOTA performance in various non-semantic speech processing tasks such as speaker recognition and synthetic speech detection.\par 

\noindent We extract representations of 1024 from TRILLsson and 768 from all the AFMs using average pooling.

\subsection{Multimodal Foundation Models}

\noindent \textbf{LanguageBind\footnote{\url{https://github.com/PKU-YuanGroup/LanguageBind}} \cite{zhu2023languagebind}:} It improves video-language (VL) frameworks by integrating multiple modalities - video, infrared, depth, and audio - with language as the central binding modality. It employs contrastive learning after freezing the language encoder and is trained on the VIDAL-10M dataset containing 10 million language-aligned pairs. 

\noindent \textbf{ImageBind\footnote{\url{https://github.com/facebookresearch/ImageBind/tree/main}} \cite{girdhar2023imagebind}:} It facilitates joint embeddings across six modalities - images, text, audio, depth, temperature, and IMU data - using image-paired data for training. It extends zero-shot capabilities across various modalities excelling in cross-modal retrieval, generation, and demonstrating robustness in few-shot learning scenarios.

\noindent \textbf{CLAP\footnote{\url{https://huggingface.co/microsoft/msclap}} \cite{elizalde2023clap}:} It aligns audio and text through natural language supervision, employing dual encoders and contrastive learning with 128,000 audio-text pairs. It achieves SOTA performance in zero-shot and supervised tasks, showcasing flexibility in class prediction and effectiveness across diverse applications without the need for labeled audio data.

\noindent We extract representations from the audio encoders of the MFMs. The dimensional size of the representations is 1024 for CLAP and ImageBind, and 768 for LanguageBind.

\section{Dimensionality Reduction Optimization}
In this section, we provide an overview of various dimensionality reduction techniques and also the methodology involved with the random selecton strategy.

\subsection{Baseline Dimensionality Reduction Techniques}

\noindent \textbf{Principal Component Analysis (PCA)~\cite{PCA}:} PCA reduces dimensionality by transforming data into a new coordinate system where the highest variances are captured by the principal components, preserving most of the dataset’s variation.

\noindent \textbf{Kernel PCA (KPCA)~\cite{KPCA}:} KPCA extends PCA using a kernel function to handle non-linear relationships by projecting data into a higher-dimensional space.

\noindent \textbf{Gaussian Random Projection (GRP)~\cite{GRP}:} GRP projects data into a lower-dimensional space using a random Gaussian matrix, preserving pairwise distances and ensuring computational efficiency.

\noindent \textbf{Singular Value Decomposition (SVD)~\cite{SVD}:} SVD decomposes a data matrix into three matrices, retaining top singular values to reduce dimensionality while preserving key features and relationships.

\begin{figure}[bt]
    \centering
    \includegraphics[scale=0.255]{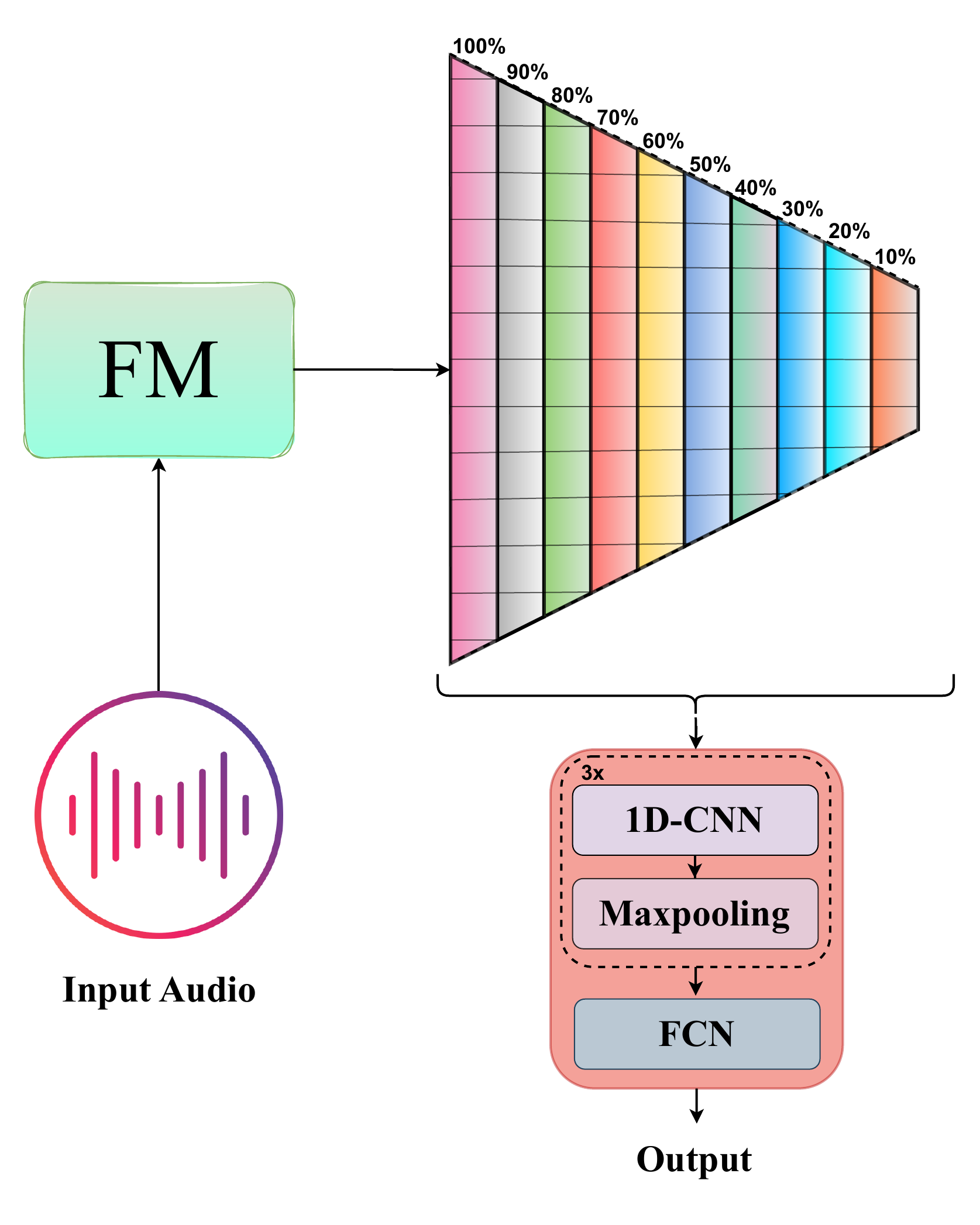}
    \caption{Modeling with a randomized selection strategy.}
    \label{fig:work_flow1}
\end{figure}

\subsection{Random Selection Strategy}
We randomly select 10\%, 20\%, 30\%, 40\%, 50\%, 60\%, 70\%, 80\%, and 90\% representation values from the representation vectors from different FMs. We train simple downstreams, Fully Connected Network (FCN) and CNN on top of the selected representation values. The modeling visualization is shown in Figure \ref{fig:work_flow1}. The FCN consisted of a dense layer with 512 neurons followed by a output layer with a single neuron with sigmoid activation function for binary classification (Fake/Non-Fake). CNN model consisted of three convolutional layers with 64, 128, and 256 filters (kernel size of 3), each followed by max-pooling and batch normalization layers. The output of the maxpooling layer is flattened and passed through FCN with a dense layer of 512 neurons and same as above FCN model.



\noindent \textbf{Modeling for Baseline Dimensionality Reduction Techniques}: We use the same downstream modeling as used for random selection. For fair comparison, we compress the representations using different dimensionality reduction techniques to the same dimension as selected percentage by random selection. For example, we randomly select 10\% representation values from a representation vector of WavLM, and the dimension of WavLM representation vector is 768. So, reduce proportionally by PCA to 10\% of 768-dimension vector. 

\section{Experiments}

\subsection{Benchmark Dataset}

We use DCASE 2023 Challenge dataset \cite{Choi_arXiv2023_01} for our experiments, which includes seven categories: dog bark, footstep, gunshot, keyboard, moving motor vehicle, rain, and sneeze/cough. The dataset consists of 5,550 authentic and 25,200 synthetic audio samples, sourced from UrbanSound8K, FSD50K, and BBC Sound Effects. All clips are 4 seconds long, mono, and sampled at 22,050 Hz. Synthetic samples were created using advanced Foley sound synthesis techniques, offering a diverse and comprehensive benchmark for EADD. The audios are resampled to 16KHz before passing to the FMs.

\begin{table}[!bt]
\centering
\begin{tabular}{c|c|c|c|c|c|c}
\toprule 
        & \multicolumn{3}{c|}{\textbf{CNN}} & \multicolumn{3}{c}{\textbf{FCN}} \\
        \hline
        & \textbf{A} & \textbf{F1} & \textbf{EER} & \textbf{A} & \textbf{F1} & \textbf{EER} \\
\midrule
\textbf{CLAP}    & 98.08 & 96.69 & 2.42   & 98.55 & 97.56 & 2.50   \\
\textbf{WavLM}   & 97.85 & 96.26 & 24.50 & \textbf{98.63} & \textbf{97.70} & 25.21  \\
\textbf{Wav2vec2} & 98.13 & 96.81 & 19.90 & 98.41 & 97.29 & 20.80 \\
\textbf{Unispeech-SAT}     & 97.82 & 96.25 & 22.35   & 98.44 & 97.37 & 22.80 \\
\textbf{TRILLsson}    & 97.74 & 96.13 & 5.21   & 98.46 & 97.38 & 5.78   \\
\textbf{LanvgaugeBind}      & \textbf{98.21} & \textbf{96.98} & \textbf{1.92}  & 98.29 & 97.12 & \textbf{2.06}    \\
\textbf{ImageBind}      & 97.90 & 96.40 & 5.37    & 98.50 & 97.46 & 6.15  \\
\textbf{MFCC}    & 98.16 & 96.88 & 18.48  & 98.33 & 97.17 & 20.34 \\
\textbf{LFCC}    & 98.07 & 96.68 & 13.14  & 98.49 & 97.42 & 15.09 \\
\bottomrule
\end{tabular}
\caption{Evaluation scores of representations from different FMs: Accuracy (A), macro-average F1 Score (F1), and Equal Error Rate (EER). Scores are in \% and represent the average of 5 folds. Higher Accuracy and F1 Score, along with a lower EER, indicate better performance. MFCC and LFCC serve as baseline features.}
\label{tab:1}
\end{table}

\begin{figure}[!ht]
    \centering
    \includegraphics[width=1\linewidth]{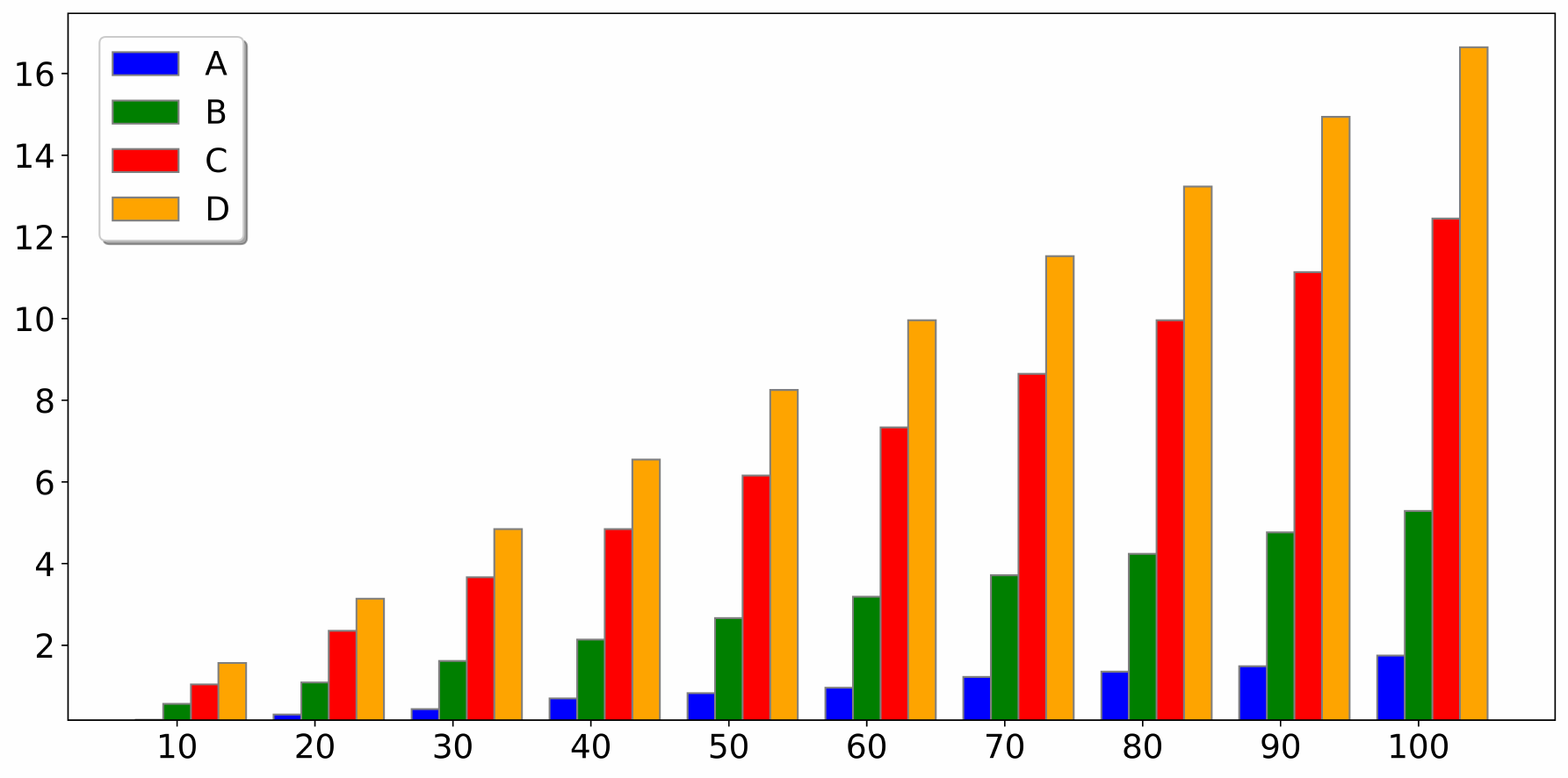}
    \caption{Barplot illustrating the impact of subset percentage (x-axis) on model parameters in millions (y-axis): \textcolor{blue}{(A) LFCC (13-dimension) }, \textcolor{green}{(B) MFCC (39-dimension)}; \textcolor{red}{(C) LanguageBind, Unispeech-SAT, wavLM, wav2vec2 (768-dimension)}; \textcolor[HTML]{ff9900}{(D) ImageBind, CLAP, TRILLsson (1024-dimension)}.}
    \label{fig:enter-label}
\end{figure}

\begin{figure}[!bt]
    \centering
    \includegraphics[scale=0.290]{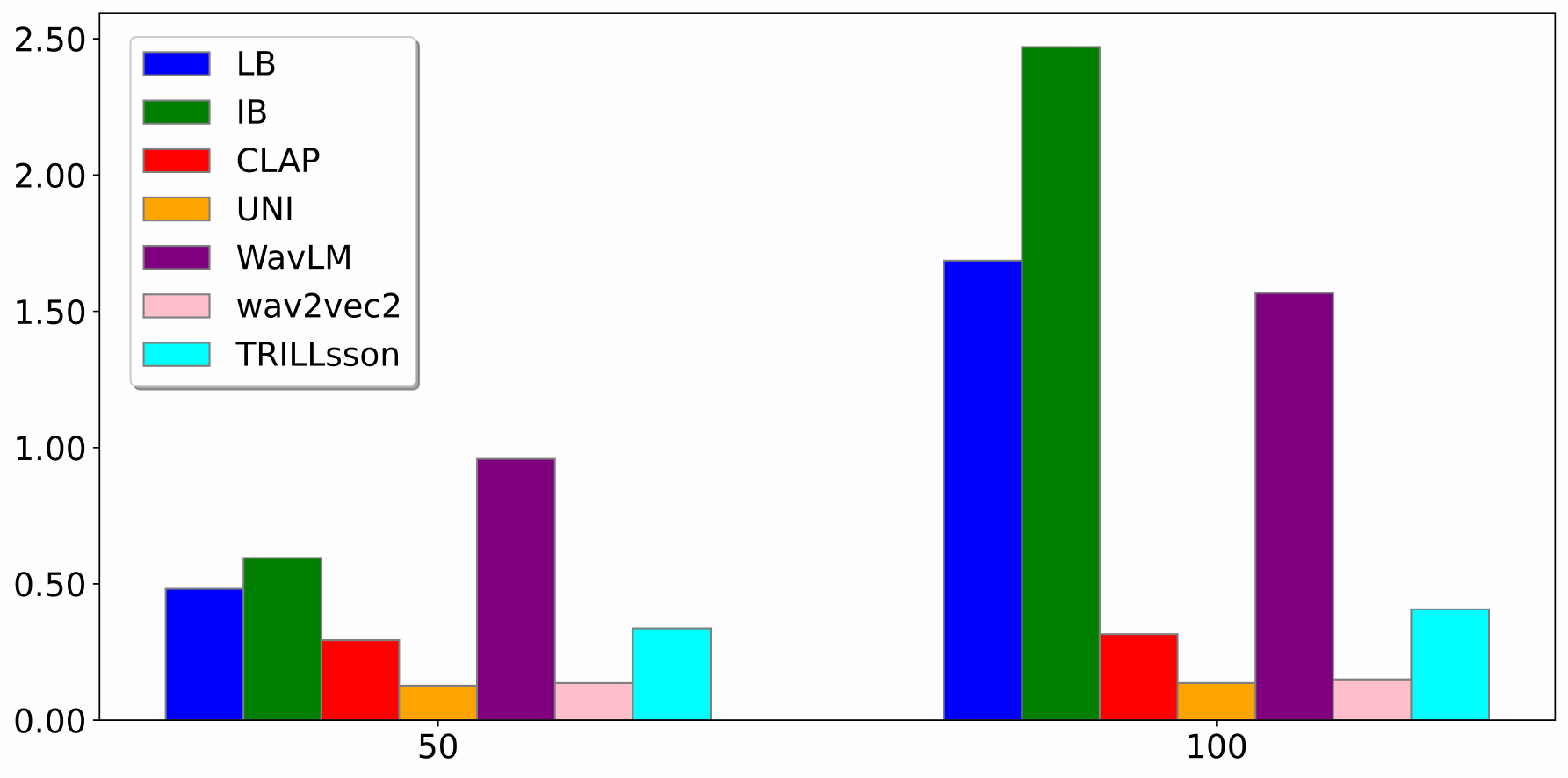}
    \caption{Inference Time Plot: x-axis represents percentage random selection of representation values; y-axis represents the inference time in seconds; LB, IB, UNI stands for LanguageBind, ImageBind, Unispeech-SAT respectively.}
    \label{fig:work_flow}
\end{figure}

\begin{figure*}[!ht]
    \centering
    \begin{minipage}{0.24\textwidth}
        \centering
        \includegraphics[width=\textwidth]{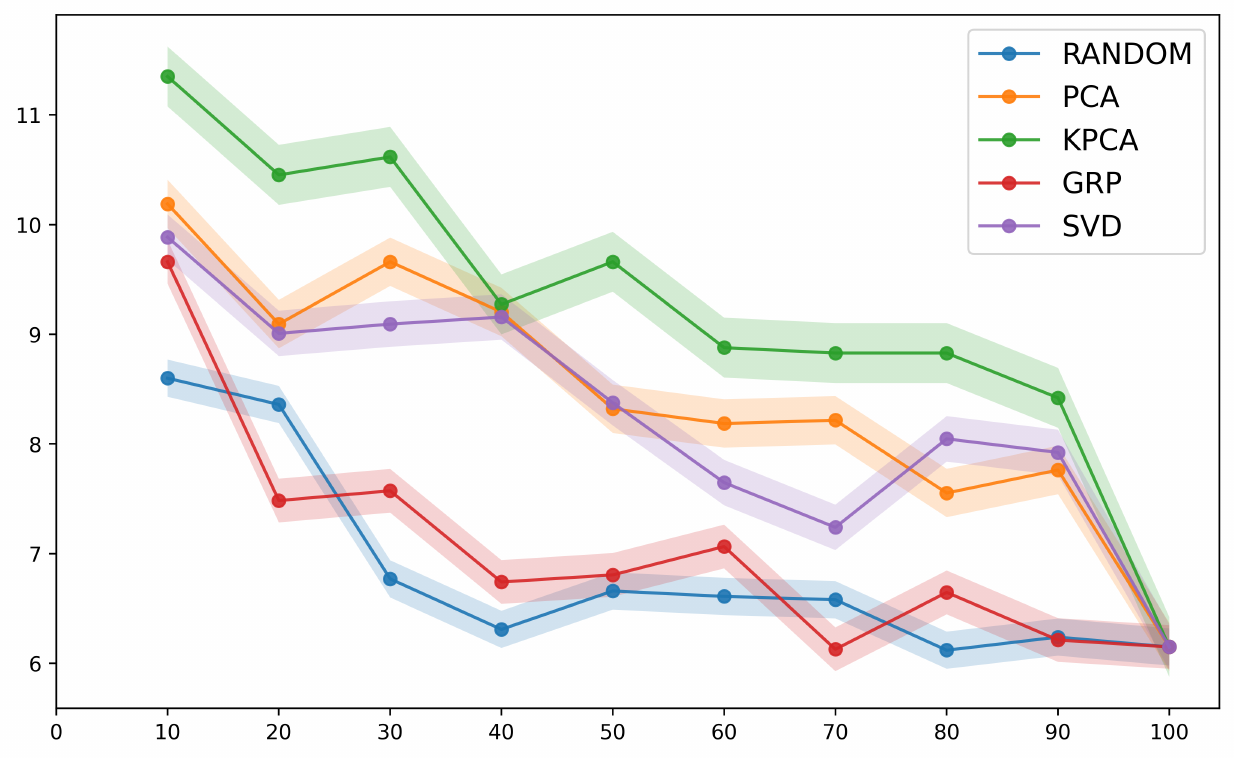} 
        \caption*{ImageBind}
    \end{minipage}\hfill
    \begin{minipage}{0.24\textwidth}
        \centering
        \includegraphics[width=\textwidth]{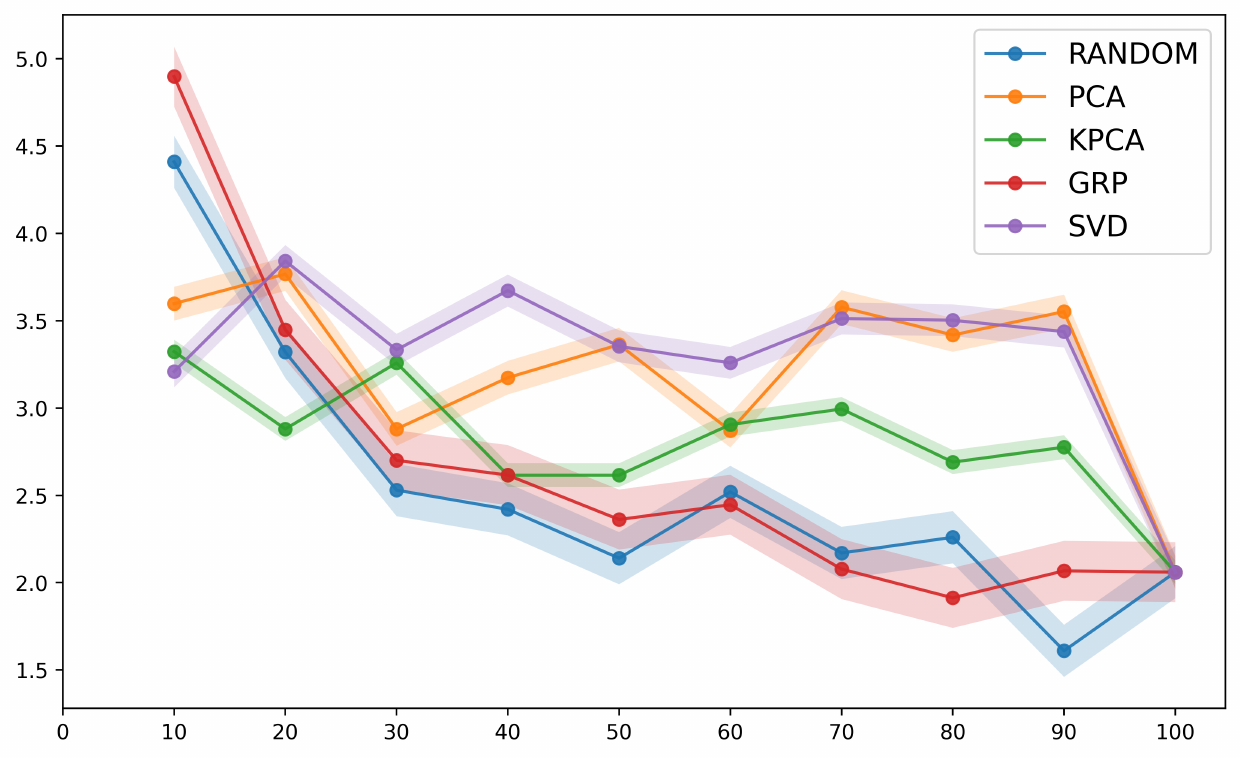} 
        \caption*{LanguageBind}
    \end{minipage}\hfill
    \begin{minipage}{0.24\textwidth}
        \centering
        \includegraphics[width=\textwidth]{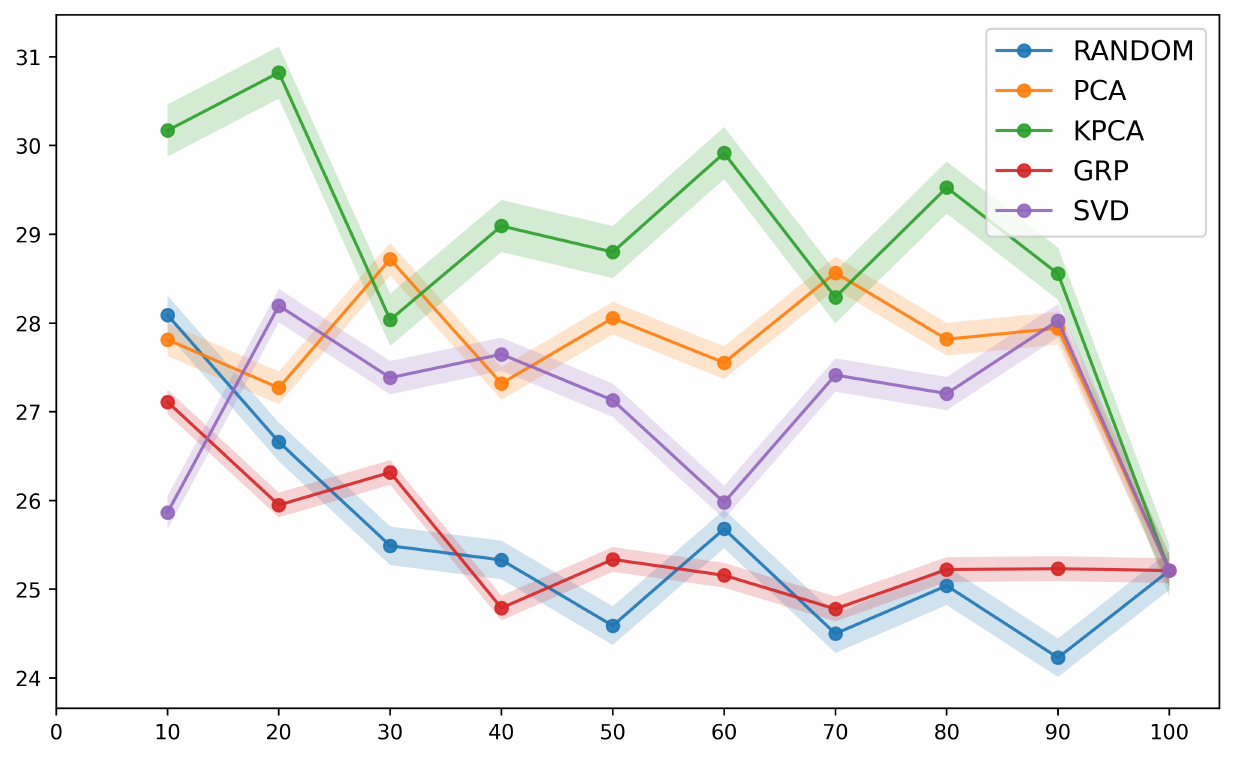} 
        \caption*{WavLM}
    \end{minipage}\hfill
    \begin{minipage}{0.24\textwidth}
        \centering
        \includegraphics[width=\textwidth]{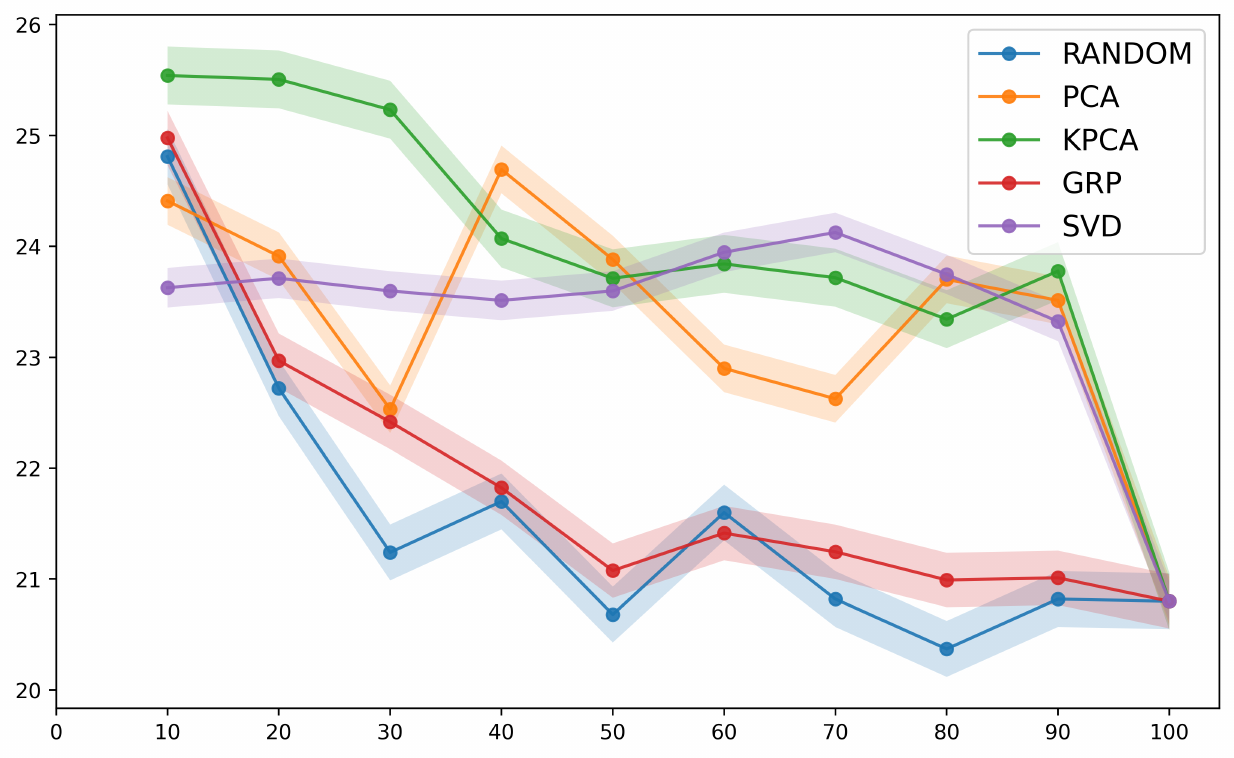} 
        \caption*{Wav2vec2}
    \end{minipage}\\[10pt] 
    \begin{minipage}{0.24\textwidth}
        \centering
        \includegraphics[width=\textwidth]{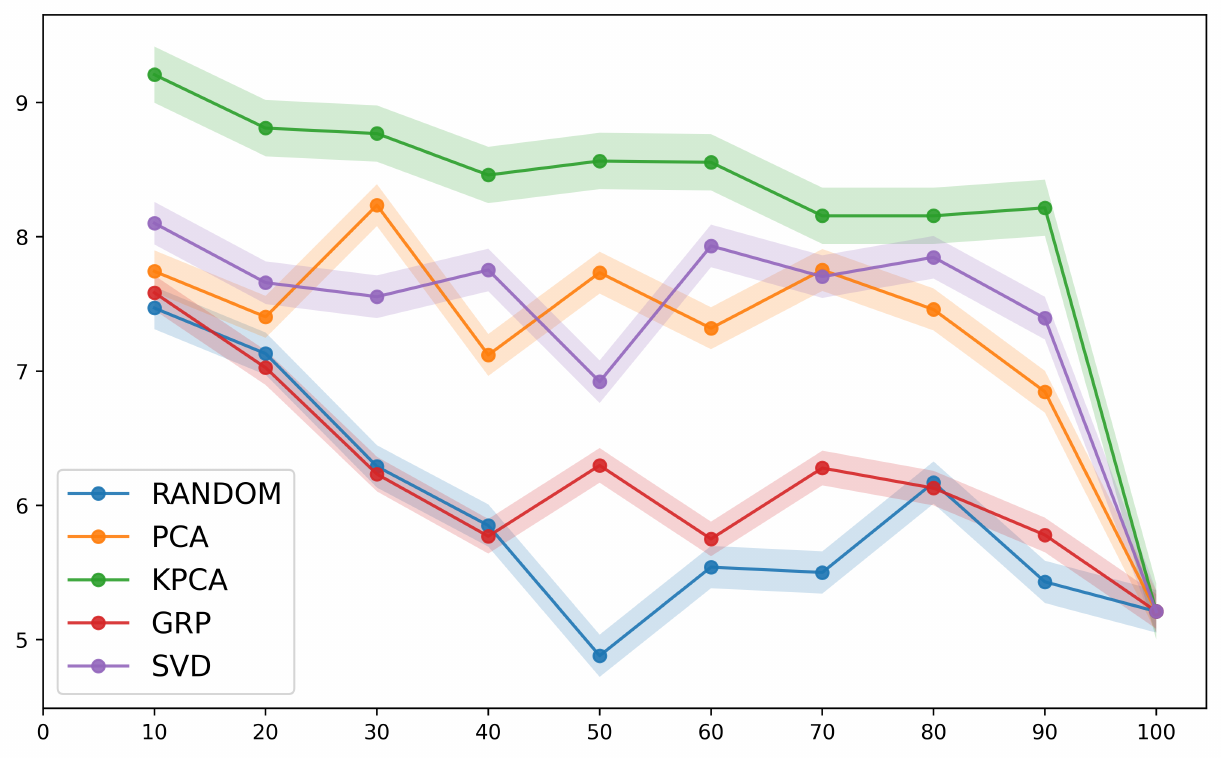} 
        \caption*{TRILLsson}
    \end{minipage}\hfill
    \begin{minipage}{0.24\textwidth}
        \centering
        \includegraphics[width=\textwidth]{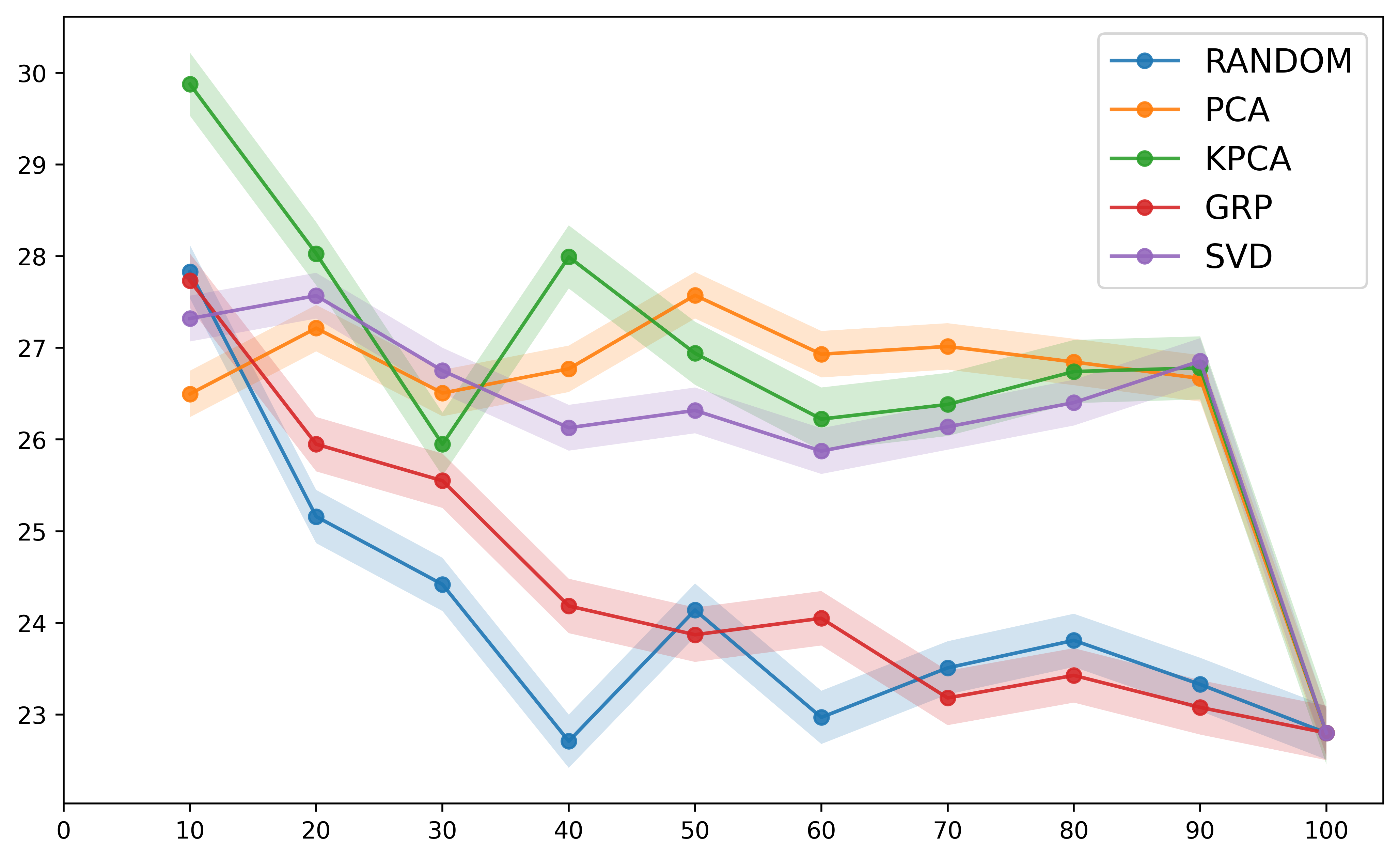} 
        \caption*{Unispeech-SAT}
    \end{minipage}\hfill
    \begin{minipage}{0.24\textwidth}
        \centering
        \includegraphics[width=\textwidth]{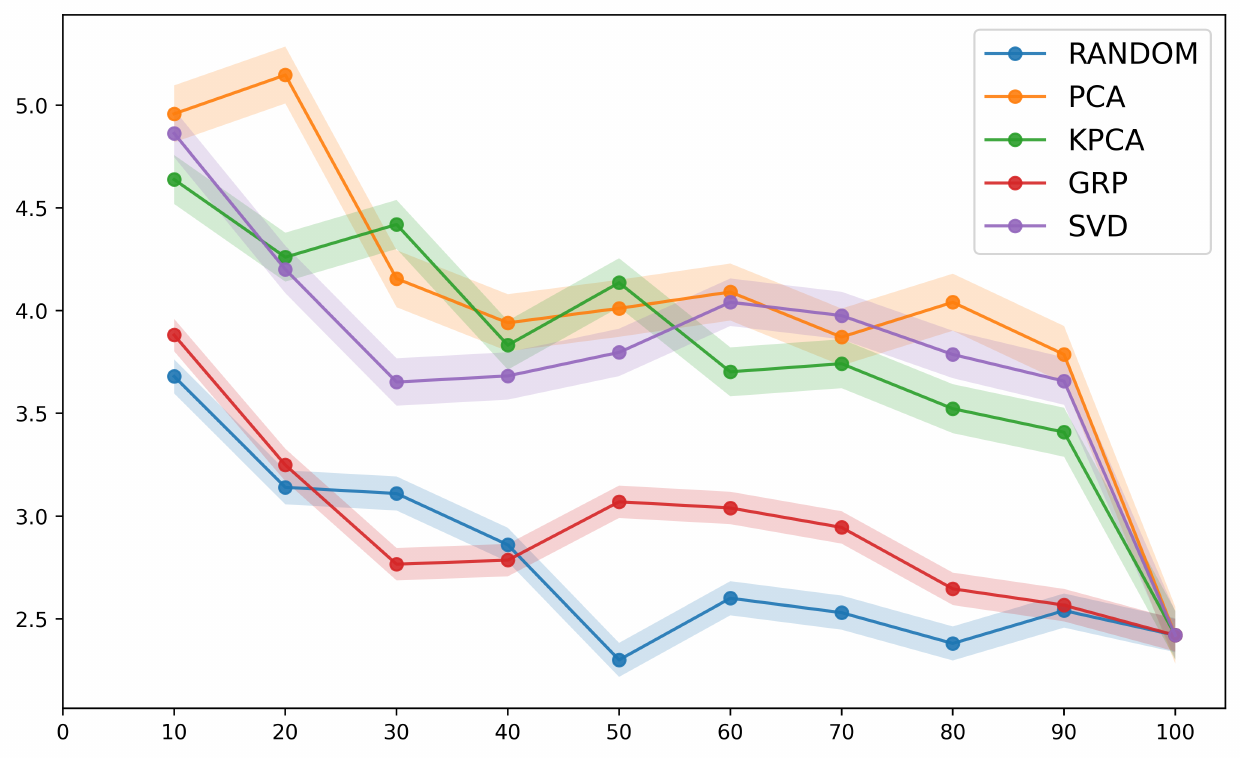} 
        \caption*{CLAP}
    \end{minipage}\hfill
    \begin{minipage}{0.24\textwidth}
        \centering
        \includegraphics[width=\textwidth]{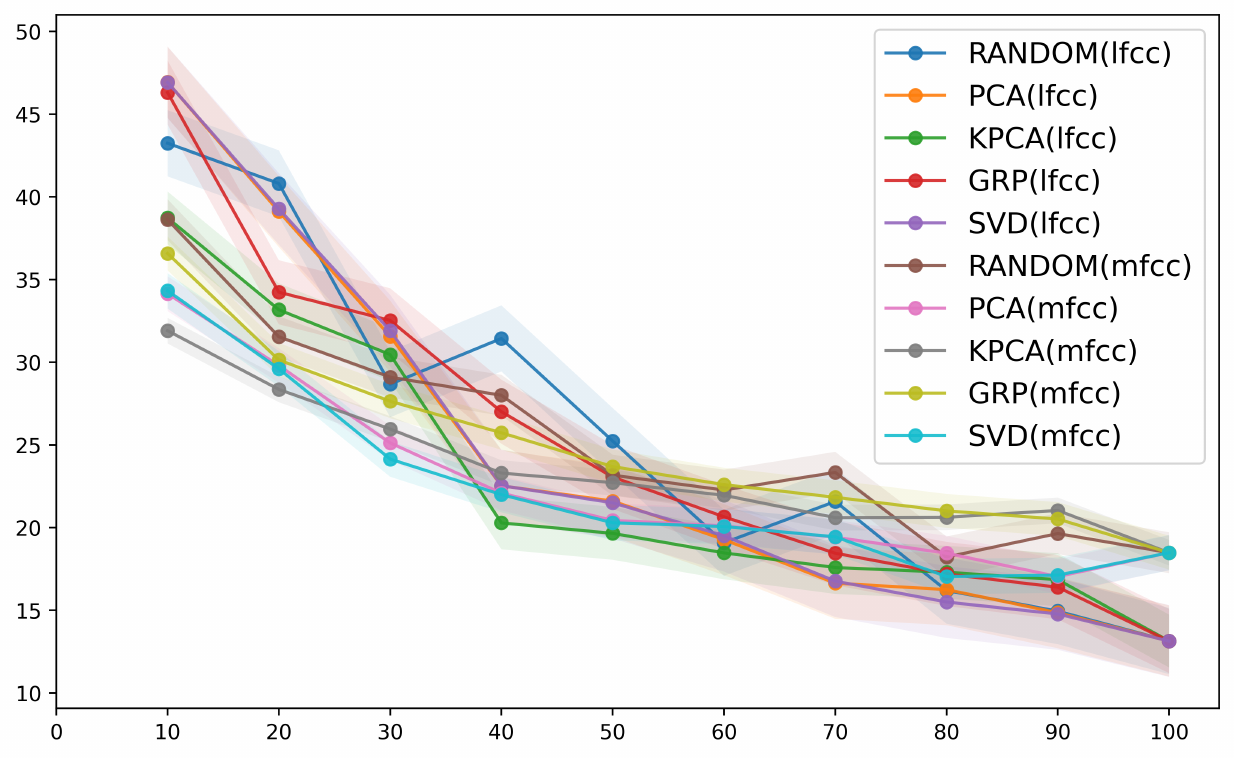} 
        \caption*{MFCC,LFCC}
    \end{minipage}
    \caption{Effect of varying proportions of representation values on EER across different FMs: y-axis in all the plots represents the EER scores in \% and x-axis in all the plots represents the percentage of representation values selected by random selection or compressed to by dimensionality reduction techniques; EER scores of random selection are given in a 5-fold manner means that 5 times random selection of representation values for a selected percentage.}
    \label{fig:subfigures}
\end{figure*}

\begin{figure*}[!ht]
    \centering
    \begin{minipage}{0.24\textwidth}
        \centering
        \includegraphics[width=\textwidth]{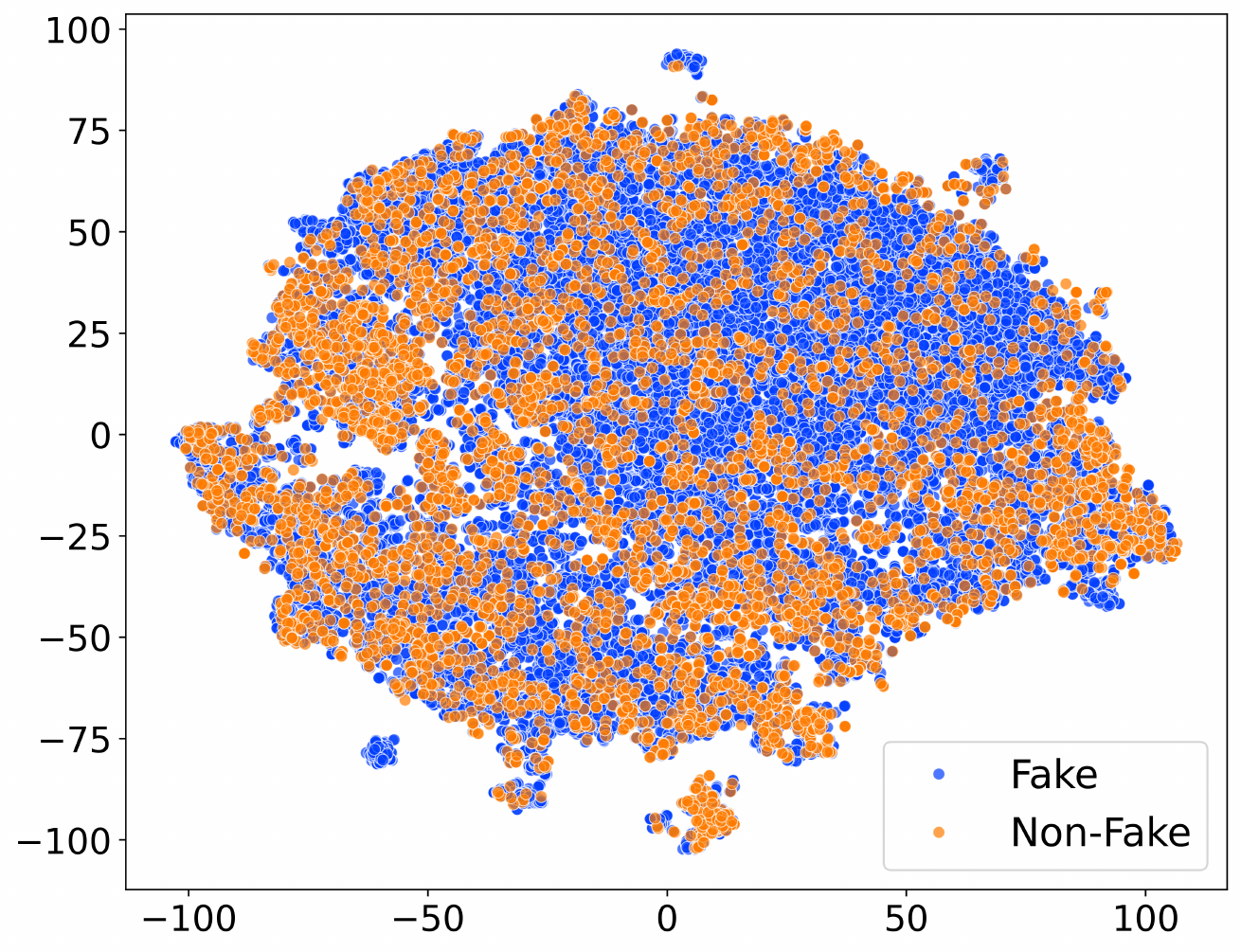} 
        \caption*{(a)}
    \end{minipage}\hfill
    \begin{minipage}{0.24\textwidth}
        \centering
        \includegraphics[width=\textwidth]{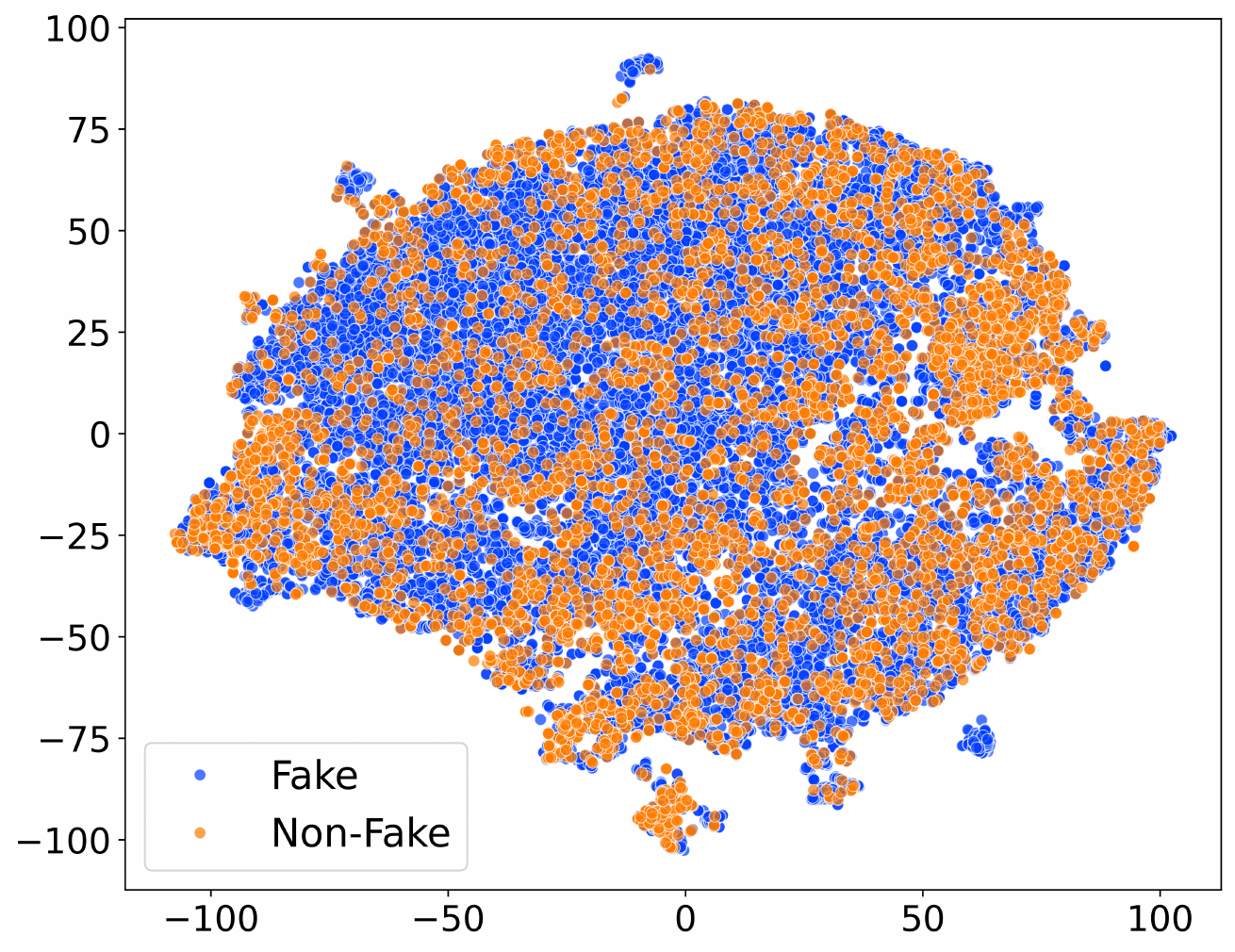} 
        \caption*{(b)}
    \end{minipage}\hfill
    \begin{minipage}{0.24\textwidth}
        \centering
        \includegraphics[width=\textwidth]{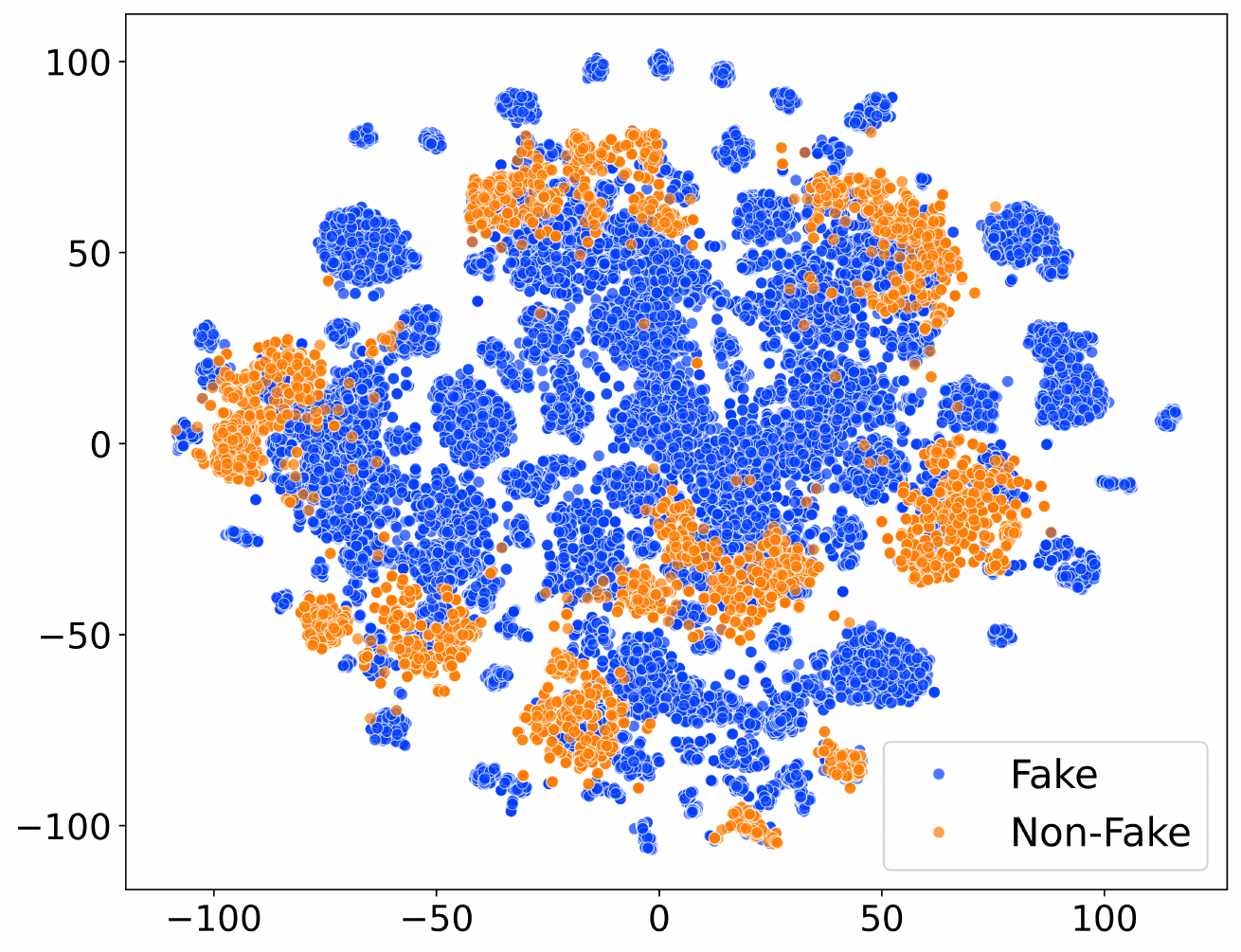} 
        \caption*{(c)}
    \end{minipage}\hfill
    \begin{minipage}{0.24\textwidth}
        \centering
        \includegraphics[width=\textwidth]{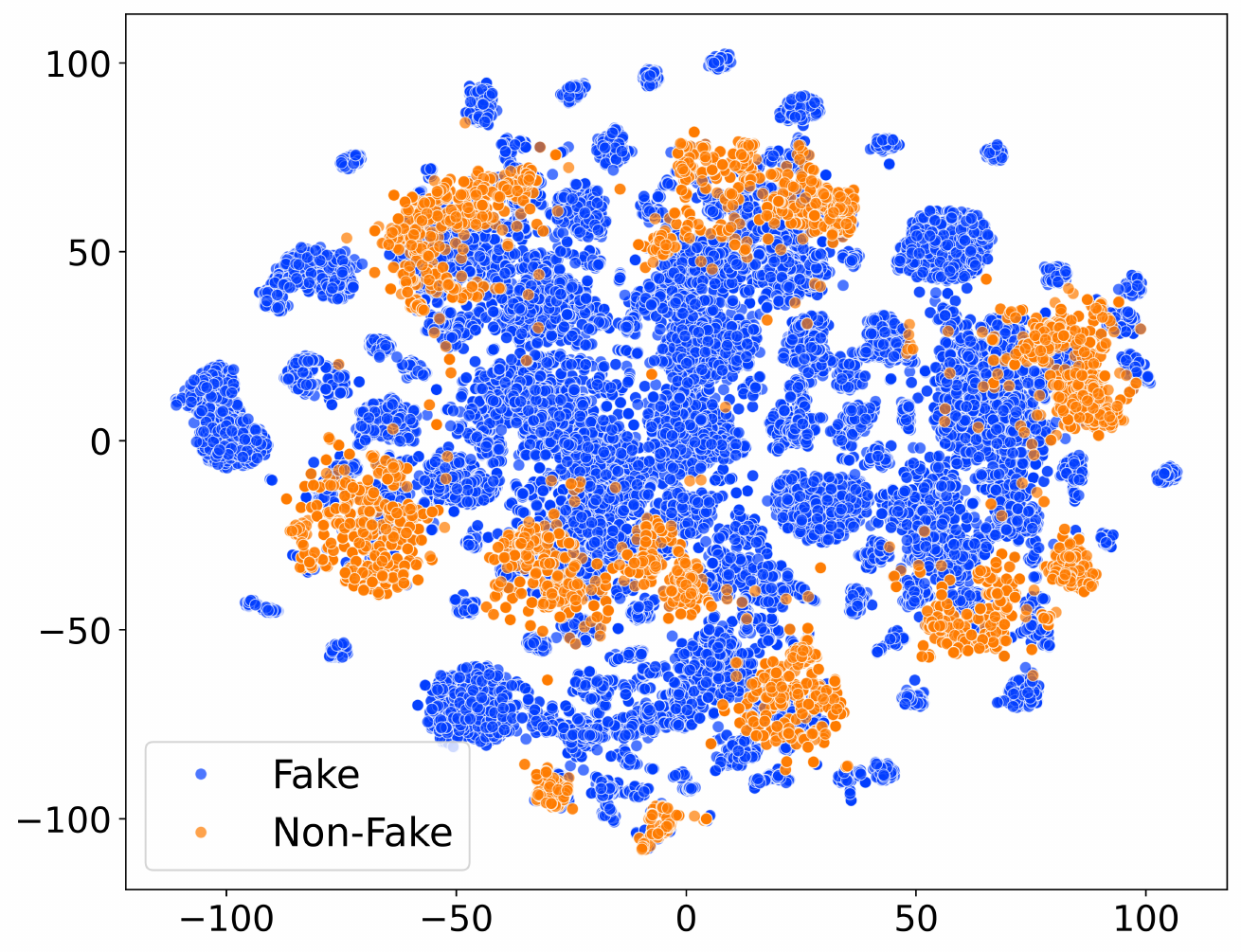} 
        \caption*{(d)}
    \end{minipage}
    \caption{t-SNE plots: (a) WavLM with 100\% representation values, (b) WavLM with 50\% representation values, (c) LanguageBind with 100\% representation values, and (d) LanguageBind with 50\% representation values.}
    \label{fig:subfigures1}
\end{figure*}

\subsection{Training Details}
All the models are trained for 50 epoch with learning rate of 1e-5. We use binary cross-entropy as the loss function and Adam as the optimizer. We also apply early stopping and dropout to prevent overfitting. We train the models in a 5-fold manner. 

\subsection{Experimental Results and Discussion}

\subsubsection{Evaluation of Foundation Models} 
First, we evaluate the full representation vector from different FMs for EADD with different downstream networks. The results are presented in Table \ref{tab:1}. We achieve better performance than previous SOTA work \cite{ouajdi2024detection} in terms of accuracy. However, through the results and also from previous research, we observe that accuracy is not the correct metric for synthetic audio detection research \cite{jung2022aasist, chetia-phukan-etal-2024-heterogeneity}, so from the next experiments onwards, we will present all the scores in terms of EER, but to show that we achieve SOTA, we compared with accuracy with the previous work as they have used accuracy as the evaluation metric. In previous SOTA work \cite{ouajdi2024detection}, the CLAP representations showed the top performance, however, through our experimental results demonstrate that the LanguageBind (LB) representations shows the topmost performance. This performance can be attributed to its pre-training with multiple modalities that is helping in better context understanding for better EADD.

\subsubsection{Dimensionality Reduction Optimization} 
Figure~\ref{fig:subfigures} shows the impact of selecting various proportions of representation values on EER. A notable trend reveals that randomly selecting 40-50\% of representation values can maintain or even enhance performance while reducing parameter counts by over half (refer to Figure~\ref{fig:enter-label}), and inference time (refer to Figure~\ref{fig:work_flow}) often outperforming SOTA dimensionality reduction techniques and achieving results comparable to or surpassing those obtained with the full 100\% set across AFMs and MFMs. This behavior remains constant in both AFMs and MFMs, which signifies that significant efficiency gains can be realized without relying solely on exhaustive representation, highlighting broader applicability across FMs compared to traditional dimensionality reduction techniques. 

Figure~\ref{fig:subfigures1} provides a t-SNE plot showcasing the distribution of representations of WavLM (100\%), WavLM (50\%), LanguageBind (100\%), and LanguageBind (50\%). The t-SNE plots reveal that reduced representation sizes (50\%) through random selection maintain a distinct separation of classes similar to the full representations (100\%), indicating that random selection of a subset of representation values does not compromise the quality of class separation which is crucial for effective classification and this is observed across t-SNE plots of MFMs as well as AFMs.

Additionally, Figure~\ref{fig:enter-label} illustrates that randomly selecting 40-50\% of representation values leads to a significant reduction in the parameter count of the downstream models. This highlights the substantial benefits of random selection and revealing how fewer parameters can achieve comparable or superior performance, thereby optimizing computational resources and downstream model efficiency.


\section{Conclusion}

In this study, we show that randomly selecting 40-50\% of representational values from representation vectors from foundation models (FMs) preserves better performance than SOTA dimensionality reduction techniques such as PCA, SVD, KPCA, and GRP. This random selection shows comparable or sometimes better performance in some FMs in comparison with the full representation vector. This random selection strategy with 40-50\% values reduces the downstream models parameters counts and inference almost to half and paves the way for efficient environmental audio deepfake detection (EADD). This behavior is observed across different AFMs as well as MFMs, showing transferability of the random selection strategy. Our study calls for future research to further investigate this behavior for making their downstream tasks efficient. 

\bibliographystyle{IEEEtran}
\bibliography{mybib}

\begin{thebibliography}{10}
\providecommand{\url}[1]{#1}
\csname url@samestyle\endcsname
\providecommand{\newblock}{\relax}
\providecommand{\bibinfo}[2]{#2}
\providecommand{\BIBentrySTDinterwordspacing}{\spaceskip=0pt\relax}
\providecommand{\BIBentryALTinterwordstretchfactor}{4}
\providecommand{\BIBentryALTinterwordspacing}{\spaceskip=\fontdimen2\font plus
\BIBentryALTinterwordstretchfactor\fontdimen3\font minus \fontdimen4\font\relax}
\providecommand{\BIBforeignlanguage}[2]{{%
\expandafter\ifx\csname l@#1\endcsname\relax
\typeout{** WARNING: IEEEtran.bst: No hyphenation pattern has been}%
\typeout{** loaded for the language `#1'. Using the pattern for}%
\typeout{** the default language instead.}%
\else
\language=\csname l@#1\endcsname
\fi
#2}}
\providecommand{\BIBdecl}{\relax}
\BIBdecl

\bibitem{DF1}
Z.~Tao, ``Deepfake generation and detection, a survey,'' \emph{Multimedia Tools and Applications}, vol.~81, pp. 6259 -- 6276, 2022.

\bibitem{DF2}
Z.~Wu, N.~W.~D. Evans, T.~H. Kinnunen, J.~Yamagishi, F.~Alegre, and H.~Li, ``Spoofing and countermeasures for speaker verification: A survey,'' \emph{Speech Commun.}, vol.~66, pp. 130--153, 2015.

\bibitem{DF3}
Y.~Zhao, J.~Yi, J.~Tao, C.~Wang, C.~Y. Zhang, T.~Wang, and Y.~Dong, ``Emofake: An initial dataset for emotion fake audio detection,'' \emph{ArXiv}, vol. abs/2211.05363, 2022.

\bibitem{DF4}
J.~Yi, Y.~Bai, J.~Tao, Z.~Tian, C.~Wang, T.~Wang, and R.~Fu, ``Half-truth: A partially fake audio detection dataset,'' \emph{ArXiv}, vol. abs/2104.03617, 2021.

\bibitem{ouajdi2024detection}
H.~Ouajdi, O.~Hadder, M.~Tailleur, M.~Lagrange, and L.~M. Heller, ``Detection of deepfake environmental audio,'' \emph{arXiv preprint arXiv:2403.17529}, 2024.

\bibitem{choi2023foley}
K.~Choi, J.~Im, L.~Heller, B.~McFee, K.~Imoto, Y.~Okamoto, M.~Lagrange, and S.~Takamichi, ``Foley sound synthesis at the dcase 2023 challenge,'' \emph{arXiv preprint arXiv:2304.12521}, 2023.

\bibitem{OLD1}
A.~K. Singh and P.~Singh, ``Detection of ai-synthesized speech using cepstral \& bispectral statistics,'' \emph{2021 IEEE 4th International Conference on Multimedia Information Processing and Retrieval (MIPR)}, pp. 412--417, 2020.

\bibitem{OLD2}
C.~Borrelli, P.~Bestagini, F.~Antonacci, A.~Sarti, and S.~Tubaro, ``Synthetic speech detection through short-term and long-term prediction traces,'' \emph{EURASIP Journal on Information Security}, vol. 2021, 2021.

\bibitem{DEEP1}
I.~Altalahin, S.~Alzu'bi, A.~A.~M. Alqudah, and A.~Mughaid, ``Unmasking the truth: A deep learning approach to detecting deepfake audio through mfcc features,'' \emph{2023 International Conference on Information Technology (ICIT)}, pp. 511--518, 2023.

\bibitem{DEEP2}
A.~Qais, A.~Rastogi, A.~Saxena, A.~Rana, and D.~Sinha, ``Deepfake audio detection with neural networks using audio features,'' \emph{2022 International Conference on Intelligent Controller and Computing for Smart Power (ICICCSP)}, pp. 1--6, 2022.

\bibitem{DEEP3}
L.~DoraM.Ballesteros, Y.~Rodr{\'i}guez-Ortega, D.~Renza, and G.~R. Arce, ``Deep4snet: deep learning for fake speech classification,'' \emph{Expert Syst. Appl.}, vol. 184, p. 115465, 2021.

\bibitem{wang2021investigating}
X.~Wang and J.~Yamagishi, ``Investigating self-supervised front ends for speech spoofing countermeasures,'' \emph{arXiv preprint arXiv:2111.07725}, 2021.

\bibitem{chetia-phukan-etal-2024-heterogeneity}
O.~Chetia~Phukan, G.~Kashyap, A.~B. Buduru, and R.~Sharma, ``Heterogeneity over homogeneity: Investigating multilingual speech pre-trained models for detecting audio deepfake,'' in \emph{Findings of the Association for Computational Linguistics: NAACL 2024}, Jun. 2024, pp. 2496--2506.

\bibitem{chen2022unispeech}
S.~Chen, Y.~Wu, C.~Wang, Z.~Chen, Z.~Chen, S.~Liu, J.~Wu, Y.~Qian, F.~Wei, J.~Li \emph{et~al.}, ``Unispeech-sat: Universal speech representation learning with speaker aware pre-training,'' in \emph{ICASSP 2022-2022 IEEE International Conference on Acoustics, Speech and Signal Processing (ICASSP)}.\hskip 1em plus 0.5em minus 0.4em\relax IEEE, 2022, pp. 6152--6156.

\bibitem{chen2022wavlm}
S.~Chen, C.~Wang, Z.~Chen, Y.~Wu, S.~Liu, Z.~Chen, J.~Li, N.~Kanda, T.~Yoshioka, X.~Xiao \emph{et~al.}, ``Wavlm: Large-scale self-supervised pre-training for full stack speech processing,'' \emph{IEEE Journal of Selected Topics in Signal Processing}, vol.~16, no.~6, pp. 1505--1518, 2022.

\bibitem{Yang2021SUPERBSP}
S.-W. Yang, P.-H. Chi, Y.-S. Chuang, C.-I. Lai, K.~Lakhotia, Y.~Y. Lin, A.~T. Liu, J.~Shi, X.~Chang, G.-T. Lin, T.~hsien Huang, W.-C. Tseng, K.~tik Lee, D.-R. Liu, Z.~Huang, S.~Dong, S.-W. Li, S.~Watanabe, A.~rahman Mohamed, and H.~yi~Lee, ``Superb: Speech processing universal performance benchmark,'' in \emph{Interspeech}, 2021.

\bibitem{baevski2020wav2vec}
A.~Baevski, Y.~Zhou, A.~Mohamed, and M.~Auli, ``wav2vec 2.0: A framework for self-supervised learning of speech representations,'' \emph{Advances in neural information processing systems}, vol.~33, pp. 12\,449--12\,460, 2020.

\bibitem{shor22_interspeech}
J.~Shor and S.~Venugopalan, ``{TRILLsson: Distilled Universal Paralinguistic Speech Representations},'' in \emph{Proc. Interspeech 2022}, 2022, pp. 356--360.

\bibitem{zhu2023languagebind}
B.~Zhu, B.~Lin, M.~Ning, Y.~Yan, J.~Cui, H.~Wang, Y.~Pang, W.~Jiang, J.~Zhang, Z.~Li \emph{et~al.}, ``Languagebind: Extending video-language pretraining to n-modality by language-based semantic alignment,'' \emph{arXiv preprint arXiv:2310.01852}, 2023.

\bibitem{girdhar2023imagebind}
R.~Girdhar, A.~El-Nouby, Z.~Liu, M.~Singh, K.~V. Alwala, A.~Joulin, and I.~Misra, ``Imagebind: One embedding space to bind them all,'' in \emph{Proceedings of the IEEE/CVF Conference on Computer Vision and Pattern Recognition}, 2023, pp. 15\,180--15\,190.

\bibitem{elizalde2023clap}
B.~Elizalde, S.~Deshmukh, M.~Al~Ismail, and H.~Wang, ``Clap learning audio concepts from natural language supervision,'' in \emph{ICASSP 2023-2023 IEEE International Conference on Acoustics, Speech and Signal Processing (ICASSP)}.\hskip 1em plus 0.5em minus 0.4em\relax IEEE, 2023, pp. 1--5.

\bibitem{PCA}
S.~Wold, K.~H. Esbensen, and P.~Geladi, ``Principal component analysis,'' \emph{Chemometrics and Intelligent Laboratory Systems}, vol.~2, pp. 37--52, 1987.

\bibitem{KPCA}
B.~Scholkopf, A.~Smola, and K.-R. M{\"u}ller, ``Kernel principal component analysis,'' in \emph{International Conference on Artificial Neural Networks}, 1997.

\bibitem{GRP}
K.~Vu, P.-L. Poirion, and L.~Liberti, ``Gaussian random projections for euclidean membership problems,'' \emph{Discrete Applied Mathematics}, vol. 253, pp. 93--102, 2019.

\bibitem{SVD}
G.~H. Golub and C.~Reinsch, ``Singular value decomposition and least squares solutions,'' \emph{Numerische Mathematik}, vol.~14, no.~5, pp. 403--420, 1970.

\bibitem{Choi_arXiv2023_01}
K.~Choi, J.~Im, L.~Heller, B.~McFee, K.~Imoto, Y.~Okamoto, M.~Lagrange, and S.~Takamichi, ``Foley sound synthesis at the dcase 2023 challenge,'' \emph{In arXiv e-prints: 2304.12521}, 2023.

\bibitem{jung2022aasist}
J.-w. Jung, H.-S. Heo, H.~Tak, H.-j. Shim, J.~S. Chung, B.-J. Lee, H.-J. Yu, and N.~Evans, ``Aasist: Audio anti-spoofing using integrated spectro-temporal graph attention networks,'' in \emph{ICASSP 2022-2022 IEEE international conference on acoustics, speech and signal processing (ICASSP)}.\hskip 1em plus 0.5em minus 0.4em\relax IEEE, 2022, pp. 6367--6371.

\end{thebibliography}

\end{document}